# Optimal Filtering of Malicious IP Sources

Fabio Soldo, Athina Markopoulou
University of California, Irvine
{fsoldo, athina}@uci.edu

Katerina Argyraki
EPFL, Switzerland
katerina.argyraki@epfl.ch

***Abstract*— How can we protect the network infrastructure from malicious traffic, such as scanning, malicious code propagation, and distributed denial-of-service (DDoS) attacks? One mechanism for blocking malicious traffic is filtering: access control lists (ACLs) can selectively block traffic based on fields of the IP header. Filters (ACLs) are already available in the routers today but are a scarce resource because they are stored in the expensive ternary content addressable memory (TCAM). In this paper, we develop, for the first time, a framework for studying filter selection as a resource allocation problem. Within this framework, we study five practical cases of source address/prefix filtering, which correspond to different attack scenarios and operator's policies. We show that filter selection optimization leads to novel variations of the multidimensional knapsack problem and we design optimal, yet computationally efficient, algorithms to solve them. We also evaluate our approach using data from *Dshield.org* and demonstrate that it brings significant benefits in practice. Our set of algorithms is a building block that can be immediately used by operators and manufacturers to block malicious traffic in a cost-efficient way.**

## I. Introduction

How can we protect our network infrastructure from malicious traffic, such as scanning, malicious code propagation, spam, and distributed denial-of-service (DDoS) attacks? These activities cause problems on a regular basis ranging from simple annoyance to severe financial, operational and political damage to companies, organizations and critical infrastructure. In recent years, they have increased in volume, sophistication, and automation, largely enabled by botnets that are used as the platform for launching these attacks.

Protecting a victim (host or network) from malicious traffic is a hard problem that requires the coordination of several complementary components, including non-technical (e.g. business and legal) and technical solutions (at the application and/or network level). Filtering support from the network is a fundamental building block in this effort. For example, the victim's ISP may install filters to react to an ongoing attack, by blocking malicious traffic before it reaches the victim. Another ISP may want to proactively identify and block the malicious traffic before it reaches and compromises vulnerable hosts in the first place. In either case, filtering is a necessary operation that must be performed within the network.

Filtering capabilities are already available at the routers today via access control lists (ACLs). ACLs allow a router to match a packet header against rules [1] and are currently used for enforcing a variety of policies, including infrastructure protection [2]. For the purpose of blocking malicious traffic, a filter is a simple ACL rule that denies access to a source IP address or prefix. To keep up with the high rates of modern routers, it is important that filtering is implemented in hardware: indeed ACLs are stored in the Ternary Content Addressable Memory (TCAM), which allows for parallel access and reduces the number of lookups per forwarded packet. However, TCAM is more expensive and consumes more space and power than conventional memory. The size and cost of TCAM puts a limit on the number of filters and this is not expected to change in the near future.[1] With thousands or tens of thousands of filters per path, an ISP alone cannot hope to block the currently witnessed attacks, not to mention attacks from multimillion-node botnets expected in the near future.

Consider the example shown in Fig.1(a): an attacker commands a large number of compromised hosts to send traffic towards a victim $V$ (say a webserver), thus exhausting the resources of $V$ and preventing it from serving its legitimate clients; the ISP of $V$ tries to protect its client from the attack, by blocking the attack at the gateway router $G$. Ideally, $G$ would like to assign a single filter to block each malicious IP source. However, there are less filters than attackers and aggregation is typically used: a single filter blocks an entire source address prefix. This has the desired effect of reducing the number of filters but also the side-effect of blocking legitimate traffic originating from that prefix. Therefore, filter selection becomes an optimization problem that tries to block as many malicious and as few legitimate sources as possible, given a certain budget on the number of filters.

In this paper, we formulate, for the first time, a general framework for studying filter selection as a resource allocation problem. To the best of our knowledge, the optimal filter selection aspect has not been explored so far, as most related work on filtering has focused on protocol and architectural aspects. Within this framework, we consider five practical source address filtering problems, depending on the attack scenario and the operator's policy and constraints. Our contributions are twofold. On the theoretical side, filter selection optimization leads to novel variations of the multidimensional knapsack problem, and we exploit the special structure of each problem to design optimal and computationally efficient algorithms. On the practical side, we provide a set of cost-

[1] A router linecard or supervisor-engine card typically supports a single TCAM chip with tens of thousands of entries. For example, the Cisco Catalyst 4500, a mid-range switch, provides a 64,000-entry TCAM to be shared among all its interfaces (48- 384). Cisco 12000, a high-end router used at the Internet core, provides 20,000 entries that operate at line-speed per linecard (up to 4 Gigabit Ethernet interfaces). The Catalyst 6500 switch can fit 16K-32K patterns and 2K-4K masks in the TCAM. Depending on how an ISP connects to its clients, each individual client can typically use only part of these ACLs, i.e. a few hundreds to a few thousands filters.

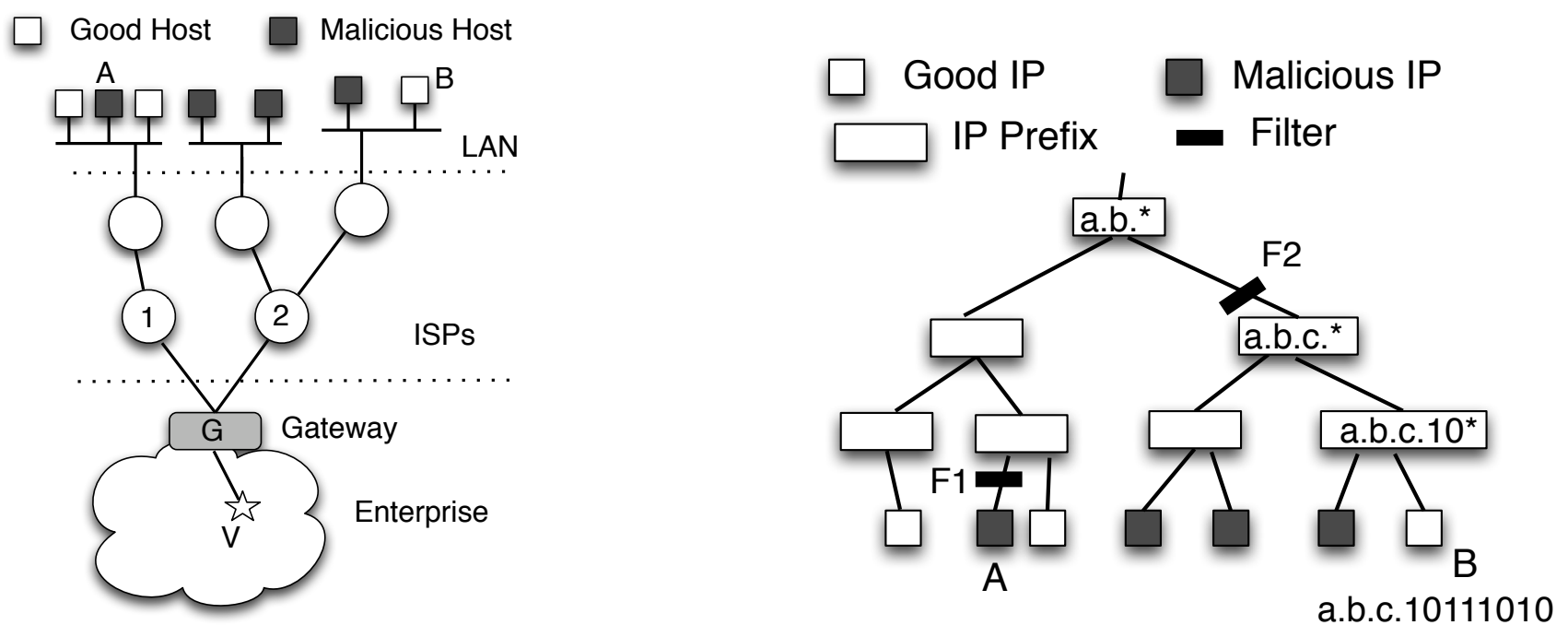

(a) Actual network. (b) Hierarchy of source IP addresses and prefixes

Fig. 1. Example of a distributed attack. Let's assume that the gateway router $G$ has only two filters available to block malicious traffic and protect the victim $V$. It uses $F1$ to block a single malicious address (A) and $F2$ to block prefix $a.b.c.*$, which contains 3 malicious sources but also one legitimate source (B). Therefore, the selection of filter $F2$ trades-off the collateral damage (blocking B) for the reduction in the number of filters (from 3 to 1).

efficient algorithms that can be used both by operators to block malicious traffic and by router manufacturers to optimize the use of their TCAM and eventually optimize the cost of the routers. We would like to emphasize that we do not propose a novel architecture for dealing with malicious traffic; instead, we optimize the use of an important mechanism that already exists on the Internet today and can be immediately used as a building block in larger defense systems, as discussed in detail in Section V-A.

The structure of the paper is as follows. In Section II-A, we formulate the general framework for studying filter selection. In Section III, we study five specific problems that correspond to different attack scenarios and operator's policies: blocking all addresses in a blacklist (BLOCK-ALL); blocking some addresses in a blacklist (BLOCK-SOME); blocking all/some addresses in a time-varying blacklist (TIME-VARYING BLOCK-ALL/SOME); blocking flows during a DDoS flooding attack to meet bandwidth constraints (FLOODING); and distributed filtering across several routers during flooding (DIST-FLOODING). For each problem, we design an optimal, yet computationally efficient, algorithm to solve it. In Section IV, we use data from Dshield.org [3] to evaluate the performance of our algorithms in realistic attack scenarios and demonstrate that they bring significant benefit in practice. In Section V, we position our work within (a) the bigger picture of defense against malicious traffic and (b) related knapsack problems. Section VI concludes the paper.

## II. Problem Formulation and Framework

### A. Definitions and Notation

Let us first define the notation used throughout the paper, also summarized in Table I.

*Source IP addresses and prefixes.* Every IPv4 address $i$ is a 32-bit sequence. Using the standard notation IP/mask we use $p/l$ to denote a prefix $p$ of length $l$ bits; $p$ and $l$ can take values $l = 0, 1, ...32$ and $p = 0, 1, ...2^l - 1$ respectively. Sometimes, for brevity, we will write simply $p$ to indicate prefix $p/l$. We write $i \in p/l$ to indicate that address $i$ is within the $2^{32-l}$ addresses covered by prefix $p/l$.

*Blacklists.* A blacklist ($\mathcal{BL}$) is a list of $N$ unique malicious source IP addresses, which send malicious traffic towards the victim. Identifying which sources are malicious and should be blocked is a difficult problem on its own right, but orthogonal to the focus of this paper. We consider that the set of malicious IP sources is accurately identified by another module (e.g. an intrusion detection system and/or historical data) in a pre-processing step and is given as input to our problem. (For a discussion of these assumptions, see Section V-A.)

An address is considered "bad" if it appears in a blacklist or "good" if it belongs to a whitelist (a set of legitimate addresses) $\mathcal{G}$, which may or may not be explicitly given. In the latter case, $\mathcal{G}$ includes all addresses that are not in $\mathcal{BL}$.

*Address Weight.* In the simplest version of the problem, an address is simply either bad or good, depending on whether it appears or not in a blacklist respectively. In a more general framework, a weight $w_i$ can be assigned to every address $i$ to indicate the importance of an address. We use $w_i \leq 0$ for every bad address $i$ to indicate the benefit from blocking it; we use $w_i \geq 0$ for every good address $i$ to indicate the collateral damage from blocking it; $w_i = 0$ indicates indifference about whether address $i$ will be blocked or not.

The weight $w_i$ can have different interpretation depending on the problem, as we will see later. First, it can capture the amount of bad/good traffic originating from an IP address and therefore the benefit/cost of blocking that address. Second, $w_i$ can express policy: e.g. depending on the amount of money gained/lost by the ISP when blocking address $i$, the operator can decide to assign large positive weights to its important customers that should not be blocked, or large negative weights to the worst attackers that must be blocked.[2]

*Filters.* In this paper, we focus on source address/prefix filtering. A filter is a simple ACL rule that specifies that all addresses in prefix $p/l$ should be blocked. $F_{max}$ denotes the maximum number of filters available in TCAM and is given as input to our problem. Notice that filter optimization is only meaningful when the number of available filters $F_{max}$ is much smaller than the number of malicious sources $N$, which is indeed the case in practice (see introduction and [1], [2]).

[2]The higher the absolute value of the weight assigned to an individual bad/good address, the higher preference to block/not block that address. If all good and bad addresses are assigned the same $w_g$ and $-w_b$ respectively, then the ratio $\frac{w_g}{w_b}$ is a parameter that the operator can tune to express how much she values low collateral damage vs. blocked malicious traffic. At the extreme, $w_i = \infty$ $(-\infty)$ indicates that address $i$ must never (always) be blocked.

| $i$ | Generic IP address |
|---|---|
| $w_i$ | Weight assigned to address $i$ |
| $\mathcal{BL}$ | Blacklist: a list of "bad" addresses |
| $N$ | Number of unique addresses in $\mathcal{BL}$ |
| $\mathcal{G}$ | Whitelist: a set of "good" addresses |
| $p/l$ (or "$p$" for short) | prefix $p$ of length $l$ bits (IP/mask notation) |
| $i \in p/l$ | address $i$ that belongs to prefix $p/l$ |
| $x_{p/l} \in \{1, 0\}$ | indicates if a filter blocks prefix $p/l$ or not |
| $g_{p/l} = \sum_{i \in p/l \cap G} w_i$ | collateral damage from filtering prefix $p/l$ |
| $b_{p/l} = \lvert \sum_{i \in p/l \cap B} w_i \rvert$ | bad traffic blocked by filtering prefix $p/l$ |
| $F_{max}$ | Maximum number of available filters |
| $z_p(F)$ (or $z_p(F, C)$) | optimal solution of subproblem considering only addresses in prefix $p$ and $F$ filters (and capacity $C$, in the case of FLOODING) |

TABLE I
NOTATION

The decision variable $x_{p/l} \in \{1, 0\}$ is 1 if a filter is assigned to block prefix $p/l$; or 0 otherwise. A filter $p/l$ blocks all $2^{32-l}$ addresses in that range. This has the desired effect of blocking all bad traffic $b_{p/l} = \lvert \sum_{i \in p/l \cap \mathcal{BL}} w_i \rvert$ and the side-effect of blocking all legitimate traffic $g_{p/l} = \sum_{i \in p/l \cap \mathcal{G}} w_i$, originating from that prefix. An effective filter should have a large benefit $b_{p/l}$ and low "collateral damage" $g_{p/l}$.

### *B. Rationale and Overview of Filtering Problems*

Given a set of malicious and legitimate sources, and a measure of their importance ($w$'s), the goal of filter selection is the construction of filtering rules, so as to minimize the impact of malicious sources on the network using the available network resources (e.g. filters and link capacity). Depending on the attack scenario, and the operator's policy and constraints, different problems may arise. E.g. the operator might want to block all malicious sources, or might tolerate to leave some unblocked; the attack might be of a low rate or a flooding attack; the operator may control one or several routers.

In the core of each filtering problem lies the following:

$$\min \sum_{p/l} \sum_{i \in p/l} w_i \cdot x_{p/l} \quad (1)$$

$$\text{s.t.} \sum_{p/l} x_{p/l} \le F_{max} \quad (2)$$

$$\sum_{p/l: i \in p/l} x_{p/l} \le 1 \quad \forall i \in \mathcal{BL} \quad (3)$$

$$x_{p/l} \in \{0, 1\} \quad \forall l = 0, ..32, p = 0, ..2^l \quad (4)$$

Eq.(1) expresses the objective to minimize the total cost for the network, which consists of two parts: the collateral damage (terms with $w_i > 0$) and the cost of leaving malicious traffic unblocked (terms with $w_i < 0$). We use the notation $\sum_{p/l}$ to denote summation over all possible prefixes $p/l$: $l = 0, ...32$, $p = 0, ...2^l - 1$. Eq.(2) expresses the constraint on the number of filters. Eq.(3) states that overlapping filters are mutually exclusive, i.e. each malicious address should be blocked at most once, otherwise filtering resources are wasted. Eq.(4) lists the decision variables $x_{p/l}$ corresponding to all possible prefixes; it is part of every optimization problem in this paper and will be omitted from now on for brevity.

Eq.(1)-(4) provide the general framework for filter selection optimization. Different filtering problems can be written as special cases within this framework, possibly with additional constraints. As we discuss in Section V-B, these are all multi-dimensional knapsack problems [4], which are in general, NP-hard. The specifics of each problem affect dramatically the complexity, which can vary from linear to NP-hard.

In this paper, we formulate five practical filtering problems, and we develop optimal, yet computationally efficient algorithms to solve them. Here, we summarize the rationale behind each problem and our main results. The exact formulation and detailed solution for each problem is provided in section III.

**[$P_1$]** ***BLOCK-ALL:*** Assume that a blacklist $\mathcal{BL}$ and a whitelist $\mathcal{G}$ is given; a weight is also associated with every good address to indicate the amount of legitimate traffic originating from that address. The limit on the number of filters is $F_{max}$. The first practical goal an operator may have is to choose a set of filters that block *all* malicious sources so as to minimize the collateral damage. We design an optimal algorithm that solves this problem at low-complexity (linearly increasing with $N$, i.e. the lowest achievable complexity for this problem).

**[$P_2$]** ***BLOCK-SOME:*** Assume that the same blacklist and whitelist are given, as in $P_1$. However, the operator may be willing to block *only some* (instead of all) malicious addresses, so as to decrease the collateral damage, at the expense of leaving some malicious traffic unblocked. She can achieve this by assigning weights $w_i > 0$ and $w_i < 0$ to good and bad addresses, respectively, to express their relative "importance". The goal of $P_2$ is to block only those subsets of malicious addresses that have the highest impact and are not co-located with important legitimate sources, so as to minimize the total cost in Eq.(1). We design an optimal, computationally efficient (linearly increasing with $N$) algorithm for this problem too.

**[$P_3$]** ***TIME-VARYING BLOCK-ALL (SOME):*** Assume that a set of blacklists $\{\mathcal{BL}_{T_0}, \mathcal{BL}_{T_1}, \ldots, \mathcal{BL}_{T_i}, \ldots\}$, and a set of whitelists $\{\mathcal{G}_{T_0}, \mathcal{G}_{T_1}, \ldots, \mathcal{G}_{T_i}, \ldots\}$ are given at different times, $T_0 < T_1 < \cdots < T_i < \ldots$; a weight is also associated with every address; the limit on the number of filters is $F_{max}$. The goal of $P_3$ is to exploit temporal correlation between blacklists at successive times and, given the solution to BLOCK-ALL(SOME) for input blacklist $\mathcal{BL}_{T_{i-1}}$, to *efficiently* update the filtering rules and construct the solution to BLOCK-ALL(SOME) with input blacklist $\mathcal{BL}_{T_i}$.

**[$P_4$]** ***FLOODING:*** In a distributed *flooding attack*, such as the one shown in Fig.1, a large number of compromised hosts send traffic to the victim with the purpose of exhausting the victim's access bandwidth. The problem is well-known and increasingly frequent and severe. Our framework can be used to optimally select filters in this case, so as to minimize the collateral damage and meet the bandwidth constraint (i.e. the total bandwidth of the unblocked traffic should not exceed the bandwidth of the flooded link, e.g. link G-V in Fig.1). The

input is the same as in $P_1$-$P_2$, and the weights capture the traffic volume originating from each IP source. We prove that the problem $P_4$ is NP-hard and we design a pseudo-polynomial algorithm that optimally solve problem $P_4$ with complexity that grows linearly with the number of sources in the blacklist and the whitelist $|\mathcal{BL}| + |\mathcal{G}|$.

**[$P_5$] *DIST-FLOODING*:** All the above problems aim at selecting filters at a single router. However, a network administrator, of an ISP or campus network, may use the filtering resources collaboratively across *several routers* to better defend against an attack. (Distributed filtering may also be enabled by the cooperation across several ISPs against a common enemy.) The question then is not only which filters to select but also on which router to place them. Here, we focus on DIST-FLOODING, which is the practical case of distributed filtering, across several routers, against a flooding attack. We prove that $P_5$ can be decomposed into several FLOODING problems, that can be solved independently and optimally one at each router.

## III. Filtering Problems and Algorithms

In this section, we give the detailed formulation of each problem and the algorithm that solves it. But first, let us define a data structure that we use to represent the problem and to develop all the subsequent algorithms.

### A. Data Structure for Representing the Problem

*Definition 1 (LCP Tree):* Given a set $\mathcal{A}$ of $N$ IP addresses, we define *the Longest Common Prefix tree* of $\mathcal{A}$, LCP($\mathcal{A}$), as the binary tree whose leaves represent the $N$ IPs and all other nodes represent all and only the longest common prefixes between any pair of IPs in $\mathcal{A}$. The prefixes are organized in the natural IP hierarchy, with shorter prefixes towards the root and longer prefixes towards the leaves, so that the prefix corresponding to a parent node includes the prefixes corresponding to its two children.

An example is shown and discussed in Fig.2.

The LCP tree can be constructed from the binary tree of all prefixes, by removing the branches that do not have malicious IPs and then by removing nodes with a single child. It reduces the storage for representing candidate prefixes by encoding those prefixes that are part of a feasible solution. The LCP tree is a variation of the binary (unibit) trie [5] but does not have nodes with a single child. We do not claim novelty in this data structure but we describe it in detail because we use it extensively in the design of the algorithms.

*Complexity:* We can build the LCP tree from $N$ malicious addresses by performing $N$ insertions in a Patricia trie [5]. To insert a string of $m$ bits, we need at most $m$ comparisons. Thus, the worst case complexity is $O(mN)$, where $m = 32$ (bits) is the constant length of an IP address.

We will make extensive use of the LCP tree in all algorithms in the rest of this section, as it provides a compact way to represent feasible solutions and to efficiently select the optimal one. Note that every node in the LCP-tree is a candidate prefix $p/l$; for brevity of notation, we will use interchangeably the notation $p/l$ and its shorter version $p$.

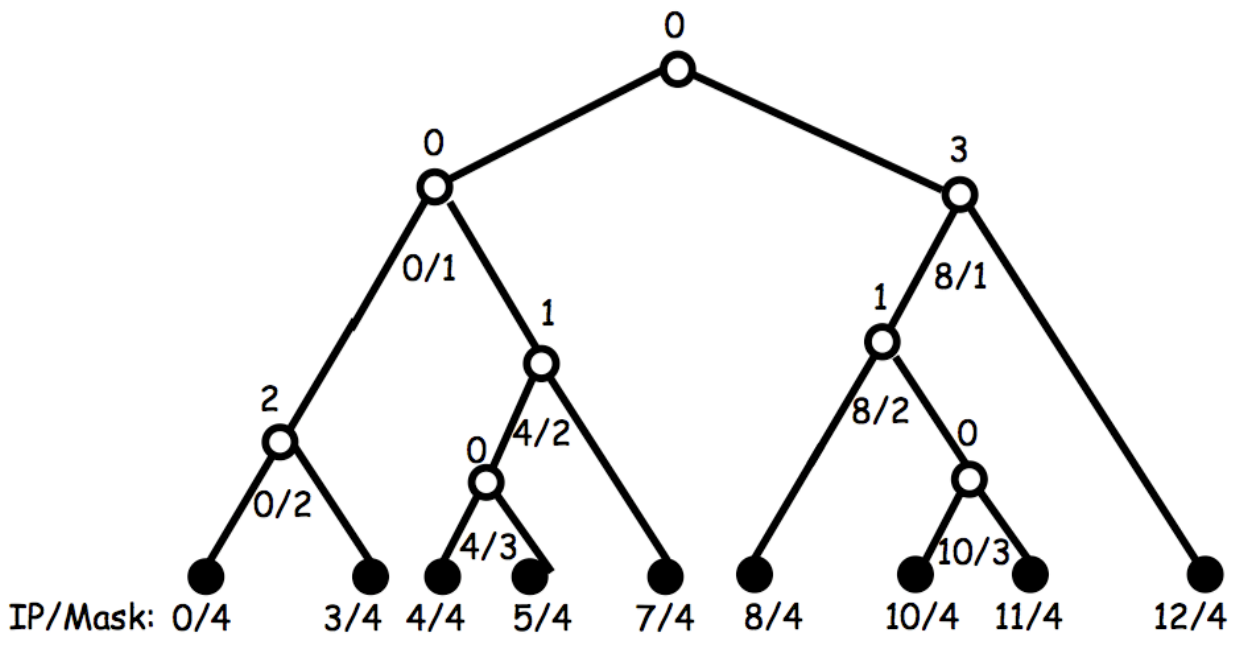

Fig. 2. **Example of LCP-tree used in BLOCK-ALL.** For ease of illustration, consider a 4-bit (instead of 32-bit) address space, i.e. from 0000 to 1111. Let $\mathcal{BL} = \{0, 3, 4, 5, 7, 810, 11, 12\}$ be the set of malicious IPs, corresponding to the leaves of the binary tree. All remaining IPs $(1, 2, 6, 9, 13, 14, 15)$ are considered legitimate and not explicitly shown. Every intermediate node represents the longest common prefix (LCP) covering all malicious sources in that subtree; it is associated with a cost measuring the additional collateral damage caused when we filtering that node, instead of filtering each of its children. E.g. the LCP of malicious addresses 0=0000 and 3=0011 is prefix 00**; if filter 00** is chosen instead of filters 0000 and 0011, collateral damage of 2 is caused, because the legitimate addresses 1 and 2 are also blocked. Choosing a set of source prefixes to filter is equivalent to choosing a set of nodes in this LCP tree. E.g. a feasible solution to BLOCK-ALL consists of prefixes $\{0/2, 4/2, 8/2, 12/4\}$ that cover all malicious IPs.

### B. BLOCK-ALL

*Goal.* Given: (i) a blacklist of malicious addresses $\mathcal{BL}$ (ii) a set of legitimate sources (iii) weights assigned to each legitimate source address, indicating the amount of traffic from that address and (iv) a limit on the number of filters $F_{max}$; select source address prefixes so as to block *all* malicious sources and minimize the collateral damage.

*Formulation.* This can be formulated within the general framework of Eq.(1)-(4) by assigning $w_i > 0$ to good addresses (the amount of legitimate traffic) and weight $w_i = 0$ to each malicious source. The goal is to minimize the total cost, which in this case is simply the total legitimate traffic blocked: $\sum_{p/l} \sum_{i \in p/l} w_i \cdot x_{p/l} = \sum_{i \in p/l \cap \mathcal{G}} w_i + 0 = g_{p/l}$. Constraint Eq.(7) enforces that every malicious source should be blocked by exactly one filter.

$$\min \sum_{p/l} g_{p/l} x_{p/l} \quad (5)$$

$$\text{s.t.} \quad \sum_{p/l} x_{p/l} \leq F_{max} \quad (6)$$

$$\sum_{p/l: i \in p/l} x_{p/l} = 1 \qquad \forall i \in \mathcal{BL} \quad (7)$$

*Characterizing an Optimal Solution.* In the algorithm, we search for solutions that can be represented as a subtree of the LCP tree structure, as described in the following:

*Proposition 3.1:* Given $\mathcal{BL}$ and $F_{max}$, there exists an optimal solution of BLOCK-ALL that can be represented as a pruned subtree of LCP-tree($\mathcal{B}L$) with: the same root, up to $F_{max}$ leaves, and non-leaf nodes having exactly two children.

*Proof:* We prove that every feasible solution of BLOCK-ALL can be reduced to another feasible solution that (i) corresponds to a subtree of LCP-tree($\mathcal{BL}$) as described in the proposition and (ii) has smaller or equal collateral damage. This is sufficient to prove the Prop.3.1 since an optimal solution is also a feasible one.

Clearly, every feasible solution of Eq. (5)-(7), $S$, can be represented as a pruned subtree of the binary tree of all possible IP prefixes, with the same root and leaves being the prefixes used as filters. Assume that $S$ uses a prefix $\tilde{p}/\tilde{l}$ which is *not* in LCP-tree($\mathcal{BL}$). Therefore, either $\tilde{p}/\tilde{l}$ does not contain any bad IPs or one of its two branches does not. In fact, if this was not the case, i.e. there is at least one bad IP in both branches, then $\tilde{p}/\tilde{l}$ would be the longest common prefix of them, and as such it would be in LCP-tree($\mathcal{BL}$).

If there are no bad IPs in prefix $\tilde{p}/\tilde{l}$, then we can safely remove the filter $\tilde{p}/\tilde{l}$, as it is not blocking any bad IPs. Similarly, if bad IPs are concentrated only in one of the two branches, then we can move the filter from $\tilde{p}/\tilde{l}$ to its child that contains all bad IP(s).

In both cases, we have a constructed a new feasible solution, with smaller (or equal) collateral damage than the original solution. Iterating this process until all prefixes are in the LCP-tree shows that any feasible solution can be transformed in a feasible solution corresponding to a subtree of LCP-tree($\mathcal{BL}$), as described in the proposition and having smaller or equal collateral damage. Therefore, also an optimal feasible solution can be transformed to that form.

Finally, we note that every node of the subtree so constructed, has two (or zero) children node. By contradiction, a set of filters which can be represented as a subtree of the LCP-tree with (at least) one node $p$ with *exactly* one child node, correspond to leaving unfiltered all bad IPs contained in the child node (prefix) of $p$ which is not selected in the subtree. [3] This violates constraint in Eq.(7), and thus correspond to a non-feasible solution of problem BLOCK-ALL. ∎

*Algorithm.* Algorithm 1, which solves BLOCK-ALL, consists of two main steps. First, we build the LCP-tree from the input blacklist. Second, in a bottom-up fashion, we compute $z_p(F) \forall p, F$, i.e. the minimum collateral damage needed to block all malicious IPs in the subtree of prefix $p$ using at most $F$ filters. Following a dynamic programming (DP) formulation, we can find the optimal allocation of filters in the subtree rooted at prefix $p$, by finding a value $n$ and assigning $F-n$ filters to the left subtree and $n$ to the right subtree, so as to minimize the collateral damage. The fact that we need to filter all malicious addresses (leaves in the LCP tree) implies that at least one filter must be assigned to the left and right subtree, i.e. $n = 1, 2..., F-1$.

[3] note that in the LCP-tree every node/prefix contain at least one bad IP.

**Algorithm 1** *Algorithm for BLOCK-ALL*

1: build LCP-tree($\mathcal{BL}$)
2: **for** all leaf nodes $leaf$ **do**
3: $z_{leaf}(F) = 0 \; \forall F \in [1, F_{max}]$
4: $X_{leaf}(F) = \{leaf\} \; \forall F \in [1, F_{max}]$
5: **end for**
6: level = level(leaf)-1
7: **while** $level \geq level(root)$ **do**
8: **for** all node $p$ such that level(p)==level **do**
9: $z_p(1) = g_p$
10: $X_p(1) = \{p\}$
11: $z_p(F) = \min_{n=1,..F-1} \left\{ z_{s_l}(F-n) + z_{s_r}(n) \right\} \forall F \in [2, F_{max}]$
12: $X_p(F) = X_{s_l}(F-n) \cup X_{s_r}(n) \forall F \in [2, F_{max}]$
13: **end for**
14: level = level - 1
15: **end while**
16: **Return** $z_{root}(F_{max}), X_{root}(F_{max})$

For every pair of sibling nodes, $s_l$ (left) and $s_r$ (right), with common parent node $p$, we have the DP recursive equation:

$$z_p(F) = \min_{n=1,\dots,F-1} \left\{ z_{s_l}(F-n) + z_{s_r}(n) \right\}, \; F > 1 \quad (8)$$

with boundary conditions for leaf and intermediate nodes:

$$z_{leaf}(F) = 0 \quad \forall F \geq 1, \quad z_p(1) = g_p \quad \forall p \quad (9)$$

Once we compute $z_p(F)$ for all prefixes in the LCP-tree, we simply read the value of the optimal solution, $z_{root}(F_{max})$. We also use the variables $X_p(F)$ to keep track of the set of prefixes used in the optimal solution. In lines (4) and (10) of Algorithm 1, $X_p(F)$ is initialized to the single prefix used. In line (12), after computing the new cost, the corresponding set of prefixes is updated: $X_p(F) = X_{s_l}(F-n) \cup X_{s_r}(n)$.

*Theorem 3.2:* Alg.1 computes the optimal solution of problem BLOCK-ALL: the prefixes that are contained in set $X_p(F)$ are the optimal $x_{p/l} = 1$ for Eq.(5)-(7).

*Proof:* Recall, $z_{root}(F_{max})$ denote the *value* of the optimal solution of BLOCK-ALL with $F_{max}$ filters (i.e. minimum amount of collateral damage), and with $X_{root}(F_{max})$ the *set of filters* selected in the optimal solution. Let $s_l$ and $s_r$ denote the two children nodes (prefixes) of *root* in the LCP-tree($\mathcal{BL}$). Finding the optimal allocation of $F_{max} > 1$ filters to block all IPs contained in *root* (possibly the all IP space), is equivalent to finding the optimal allocation of $x \geq 1$ filters to block all IPs in $s_l$, and $y \geq 1$ prefixes for bad IPs in $s_r$, such that $x + y = F_{max}$. This is because prefixes $s_l$, and $s_r$ jointly contain *all* bad IPs. Moreover, both $s_l$ and $s_r$ contains at least one bad IP. Thus, at least one filter must be assigned to each of them. If $F_{max} = 1$, i.e. there is only one filter available, the only feasible solution is to select *root* as the prefix to filter out. The same argument recursively applies to descendant nodes, until either we reach a leaf node, or we have only one filter available. In these cases, the problem is trivially solved by condition in Eq.(9). ∎

*Complexity.* Computing Eq.(8) for every node $p$ and for every $F \in [1, F_{max}-1]$ involves $N(F_{max}-1)$ subproblems, one for every pair ($p$, $F$) with complexity $F_{max}-1$ each.

$z_p(F)$ in Eq.(8) requires only the optimal solution at the sibling nodes, $z(s_l, F-n), z(s_r, n)$. Thus, proceeding from the leaves to the root, we can compute the optimal solution in $N(F_{max}-1)^2$. This simple bound can be made tighter observing that, at every node in the LCP-tree we do not need to compute $z_p(F)$ for all values $F \leq F_{max}$, but only for $F \leq \min\{|leaves(p)|, F_{max}\}$, where $|leaves(p)|$ is the number of the leaves under prefix $p$ in the LCP tree. Moreover, the complexity of computing every single entry $z_p(F)$ is obviously $F$. Thus, the overall number of operations needed equals,

$$\sum_{i \in \text{Node}} \frac{\Delta_i(\Delta_i+1)}{2} \tag{10}$$

where $\Delta_i = \min\{F_{max}, |leaves(i)|\}$. Let $L_i$ denote the level of node $i$ in the LCP-tree, with the convention that we assign $L=0$ to the root node. Per every node, such that $L_i \leq \lfloor \log\left(\frac{N}{F_{max}}\right) \rfloor$, $\Delta_i = F_{max}$; otherwise, $\Delta_i = |leaves(i)| \leq \frac{N}{2^{L_i}}$, since LCP-tree is a binary tree. Thus, we have

$$\begin{aligned}
&\sum_{L=0}^{\lceil \log N \rceil} \sum_{\substack{i \in \text{Node} \\ level(i)=L}} \frac{\Delta_i(\Delta_i+1)}{2} = \\
&= \sum_{L=0}^{\lfloor \log\left(\frac{N}{F_{max}}\right) \rfloor} \sum_{\substack{i \in \text{Node} \\ level(i)=L}} \frac{F_{max}(F_{max}+1)}{2} + \\
&+ \sum_{L=\lfloor \log\left(\frac{N}{F_{max}}\right) \rfloor+1}^{\lceil \log N \rceil} \sum_{\substack{i \in \text{Node} \\ level(i)=L}} \frac{N}{2^{L_i}}\left(\frac{N}{2^{L_i}}+1\right) = \\
&= \sum_{L=0}^{\lfloor \log\left(\frac{N}{F_{max}}\right) \rfloor} \sum_{\substack{i \in \text{Node} \\ level(i)=L}} \frac{F_{max}(F_{max}+1)}{2} + \\
&+ \sum_{L=\lfloor \log\left(\frac{N}{F_{max}}\right) \rfloor+1}^{\lceil \log N \rceil} \sum_{\substack{i \in \text{Node} \\ level(i)=L}} \frac{N^2}{2^{2L_i}} + \frac{N}{2^{L_i}} \\
&\leq (2^{\log\left(\frac{N}{F_{max}}\right)+1} - 1)\frac{F_{max}(F_{max}+1)}{2} + \\
&+ \frac{F_{max}}{2}\left(\frac{F_{max}}{2}+1\right) \\
&\leq N\frac{(F_{max}+1)}{2} + \frac{F_{max}}{2}\left(\frac{F_{max}}{2}+1\right)
\end{aligned} \tag{11}$$

where Eq.(11) uses the fact that if $0 \leq n_0 < n_1$, then $\sum_{h=n_0}^{n_1} \frac{1}{2^h} \leq \frac{1}{2^{n_0-1}}$.

Using this observation, the computation can be done in $O(NF_{max})$, which is essentially $O(N)$, since $F_{max} << N$ and $F_{max}$ does not depend on $N$ but only on the TCAM size. Thus, the time complexity increases linearly with the number of malicious IPs $N$. This is the lowest achievable complexity, within a constant factor, since we need to read all $N$ malicious IPs at least once.

### C. BLOCK-SOME

*Goal.* Given: (i) a blacklist of malicious addresses (ii) a set of legitimate sources (iii) weights assigned to all addresses, which express relative importance and (iv) a limit on the number of filters $F_{max}$; select some source address prefixes to block so as to minimize the total cost, including the cost of collateral damage and the benefit of blocking malicious addresses.

*Formulation.* This can be formulated within the general framework of Eq.(1)-(4), by assigning to good and bad addresses weights $w_i > 0$ and $w_i < 0$ respectively, to express their relative importance. The goal is to minimize the total cost, as in Eq.(1), which in this case includes both collateral damage $g_{p/l}$ and unfiltered malicious traffic $b_{p/l}$.

$$\min \sum_{p/l} \left(g_{p/l} - b_{p/l}\right) x_{p/l} \tag{12}$$

$$\text{s.t.} \sum_{p/l} x_{p/l} \leq F_{max} \tag{13}$$

$$\sum_{p/l: i \in p/l} x_{p/l} \leq 1 \qquad \forall i \in \mathcal{BL} \tag{14}$$

Another difference from BLOCK-ALL is Eq.(14), which dictates that every malicious source must be covered *at most by one prefix*, but does not necessarily have to be covered.

*Characterizing an Optimal Solution.* We can leverage again the structure of the LCP tree to characterize feasible and optimal solutions, with a proposition similar to Prop.3.1. The difference from BLOCK-ALL is that, because some bad IPs can remain unfiltered, the pruned subtree corresponding to a feasible solution can now have nodes with a single descendant.

*Proposition 3.3:* Given $\mathcal{BL}$ and $F_{max}$, there exists an optimal solution of BLOCK-SOME that can be represented as a pruned subtree of LCP-tree($\mathcal{BL}$) with: the same root, up to $F_{max}$ leaves.

*Proof:* In Prop.3.1 we proved that any solution of Eq.5-6 can be reduced to a (pruned) subtree of the LCP-tree with at most $F_{max}$ leaves. Moreover, we note that constraint in Eq.(14), which imposes the use of non-overlapping prefixes, is automatically imposed considering the *leaves* of the pruned subtree as the selected filter. This prove that any feasible solution of BLOCK-SOME can be transformed in a pruned subtree of the LCP-tree with at most $F_{max}$ leaves. And thus, can an optimal solution. ■

*Algorithm.* The algorithm is similar to Algorithm 1 in that it uses the LCP-tree and a similar DP approach. The difference is that not all addresses need to be covered and, at each step, we can assign $n=0$ filters to the left or right subtree, i.e. in line (11) of Algorithm 1: $n = 0, 1..., F$. We can recursively compute the optimal solution as before:

$$z_p(F) = \min_{n=0,\dots,F} \left\{ z_{s_l}(F-n) + z_{s_r}(n) \right\} \tag{15}$$

with boundary conditions for intermediate ($p$) and leaf nodes:

$$z_p(0) = 0 \quad \forall\, p \tag{16}$$

$$z_p(1) = \min\left\{g_p - b_p,\, \min_{n=0,1}\left\{z_{s_l}(1-n) + z_{s_r}(n)\right\}\right\} \tag{17}$$

$$z_{leaf}(F) = -b_{leaf} \quad \forall F \geq 1 \tag{18}$$

*Complexity.* The analysis of BLOCK-ALL can be applied to this algorithm as well. The complexity turns out to be the same, i.e. linearly increasing in $N$ as well.

*BLOCK-ALL vs. BLOCK-SOME.* There is an interesting connection between the two problems. The latter can be regarded as an automatic way to select the best subset from $\mathcal{BL}$, in terms of the weights $w_i$, and run BLOCK-ALL only on that subset. The advantage is that we do not need to search for the optimal subset, which is automatically given in the final solution. In the extreme case that much more importance is given to the bad rather than the good addresses, BLOCK-SOME degenerates to BLOCK-ALL.

### D. *TIME-VARYING BLOCK-ALL(SOME)*

So far, we have considered the static problem of filtering a *fixed* set of source IP addresses. However, malicious source IPs appear/disappear/reappear in a blacklist over time [9]. In this section, we consider the problem of filtering a *dynamic* set of source IPs, i.e., varying over time. This is equivalent to considering different blacklists, one at every time an IP is inserted or deleted from the blacklist. Let us denote $\{\mathcal{BL}_{T_0}, \mathcal{BL}_{T_1}, \ldots, \mathcal{BL}_{T_i}, \ldots\}$ the set of different blacklists as sampled at time $T_0 < T_1 < \cdots < T_i < \ldots$, when a new IP is inserted in the blacklist or an old one is removed. The trivial approach to the dynamic BLOCK-ALL problem is to run Alg.1 from scratch at every time instance. As noted the computational complexity of Alg.1 is low: it grows linearly with the number of IP addresses in the blacklist, $N$. However, if the overlap between two successive blacklists is large enough, we can exploit the correlation between them to construct a more efficient scheme, which updates filters as needed, while leaving most of them unchanged. More formally, consider the following problem:

*Goal.* Given a set of blacklists $\{\mathcal{BL}_{T_0}, \mathcal{BL}_{T_1}, \ldots, \mathcal{BL}_{T_i}, \ldots\}$ collected at different times, $T_0 < T_1 < \cdots < T_i < \ldots$, and $F_{max}$ filters, find the set of filtering rules $\{\mathcal{S}_{T_0}, \mathcal{S}_{T_1}, \ldots, \mathcal{S}_{T_i}, \ldots\}$ at every time such that, $\forall i = 0, 1, \ldots$ $\mathcal{S}_{T_\rangle}$ solves BLOCK-ALL(SOME) for input blacklist $\mathcal{BL}_{T_i}$.

*Algorithm.* As mentioned above, if there is no or low overlap between successive blacklists, the obvious solution to this problem is to run the BLOCK-ALL algorithm at every time a new blacklist is provided. Otherwise, if only few IPs are inserted/removed from a blacklist to the successive one, we can update all and only the filters affected by that change. For example, consider two blacklists, $\mathcal{BL}_{T_{i-1}}, \mathcal{BL}_{T_i}$, which differ only in a single new IP inserted in $\mathcal{BL}_{T_i}$. Assume that $\mathcal{S}_{T_{i-1}}$, the solution to the BLOCK-ALL problem with blacklist $\mathcal{BL}_{T_{i-1}}$, has already been computed. We want to find an efficient algorithm that computes $\mathcal{S}_{T_i}$.

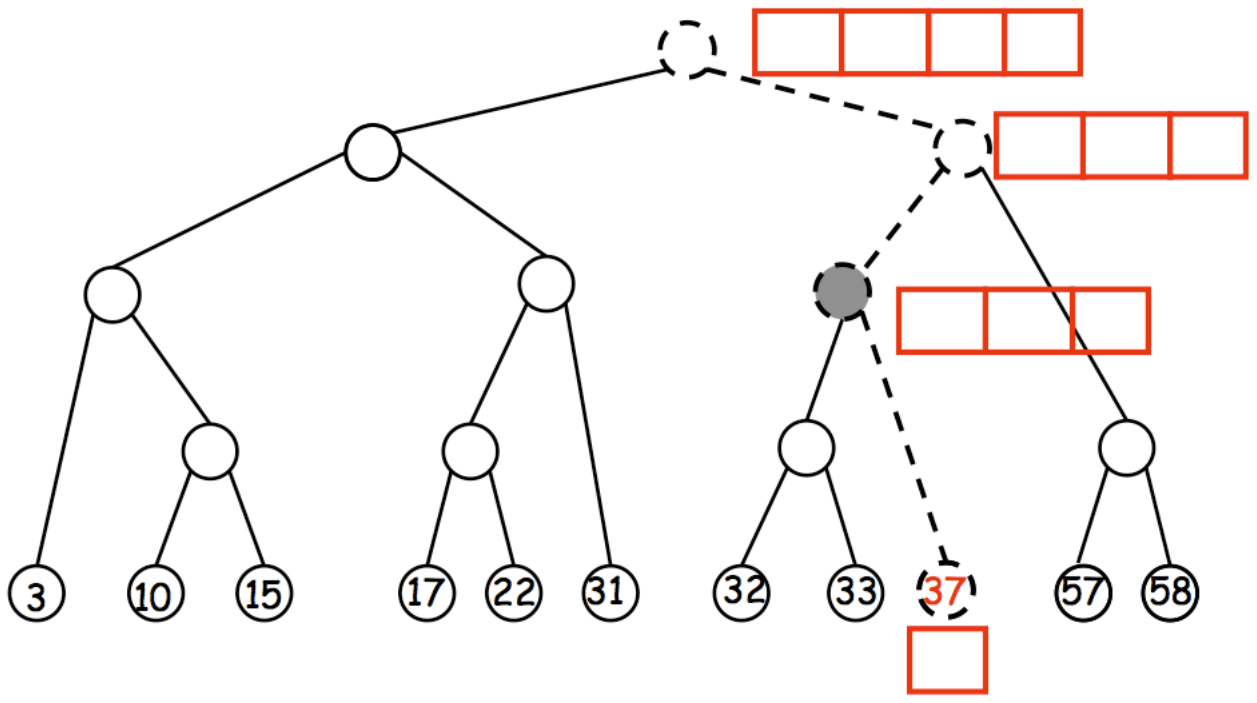

Fig. 3. As an example, assume having a 6-bits IP space, instead of the usual 32 bits. A new IP, corresponding to 37 in decimal notation, is inserted in the blacklist made up of IPs: 3,10,15,17,22,31,32,33,57,58. Its insertion requires that all and only its predecessor nodes in the LCP-tree are updated according to Eq.(8) (or Eq.(15) if we are running BLOCK-SOME). Moreover, a new node, in gray, is created to denote the longest common prefix between 37 and 32 (or 33). Note that, all other nodes corresponding to the longest common prefixes between 37 and other IPs in the blacklist, is already in the initial LCP-tree.

Basically, there are to two separate cases depending on whether or not the new IP is covered by some prefix which is already filtered in $\mathcal{S}_{T_{i-1}}$. If this is the case, no further action is needed, and $\mathcal{S}_{T_i} = \mathcal{S}_{T_{i-1}}$. Otherwise, we need to modify the filters to also cover the new IP. An efficient way to do so, is illustrated in Fig.3. When a new IP appears in the blacklist, only one intermediate node needs to be added to the LCP-tree: the one corresponding to the longest common prefix between the new node the and its "closest" IP already in the blacklist (gray node in Fig.3). As learnt from the previous sections, an optimal allocation of $f$ filters at prefix $p/l$, depends only on how these $f$ filters are allocated to the children nodes of $p$ in the LCP-tree. Thus, the insertion of a new IP in the blacklist requires only the re-computation of $z_p(f)$ and $X_p(f) \forall f$, through Eq.(8), for all and only the predecessors of the new node in the LCP-tree (nodes along the dashed path in Fig.3). Multiple insertions can be handled by iterating the above procedures for every insertion. Handling removal operations (i.e. IPs that are removed from a blacklist) is similar: when removing an IP, we also remove its parent node, since it stops being the longest common prefix of two IPs, and we update all other predecessor nodes according to Eq.(8).

We note that since any LCP-tree is also a binary tree, there are at most $\log(N)$ predecessors of any leaf node, thus the above procedure requires $O(\log(N)F_{max})$ operations. This is a more efficient update scheme, than running Alg.1 from scratch, if and only if the number of insert/remove operations that need to be performed to obtain $\mathcal{BL}_{T_i}$ from the previous blacklist, $\mathcal{BL}_{T_{i-1}}$, is less than $\frac{N}{\log N}$. Otherwise it is less expensive to simply run Alg.1 with input the new blacklist, $\mathcal{BL}_{T_i}$.

Finally, we note that we can use the same approach to solve the dynamic BLOCK-SOME problem. In that case as well, arrivals/departures of malicious addresses from the blacklist can be handled by insertions/deletions in/from the LCP-tree; the filters should be updated accordingly so that they provide an optimal solution to the static BLOCK-SOME problem for the input blacklist at every time.

### *E. FLOODING*

*Goal.* Given: (i) a blacklist of malicious addresses (ii) a set of legitimate sources (ii) *the amount of traffic* that each generates (iii) a limit on the number of filters $F_{max}$ and (iv) a constraint on the link *capacity* (bandwidth) $C$; select some source address prefixes to block so as to minimize the collateral damage and make the total traffic fit within the link capacity.

*Formulation.*

$$\min \sum_{p/l} g_{p/l} x_{p/l} \tag{19}$$

$$\text{s.t.} \sum_{p/l} x_{p/l} \leq F_{max} \tag{20}$$

$$\sum_{p/l} \Big(g_{p/l} + b_{p/l}\Big)(1 - x_{p/l}) \leq C \tag{21}$$

$$\sum_{p/l: i \in p/l} x_{p/l} \leq 1 \quad \forall i \in \mathcal{BL} \tag{22}$$

where, $g_{p/l}$ and $b_{p/l}$ denote the amount of good bad traffic from prefix $p/l$, respectively. Eq.(22) indicates that we are interested in blocking some, not all, malicious sources, and that we should not use overlapping prefixes. Before the attack, the total good traffic $t_0 = \sum_{p/l} \Big(g_{p/l} + b_{p/l}\Big)$ could fit within the capacity; after flooding, the total traffic exceeds the capacity. Eq.(21) says that the total traffic that remains unblocked after filtering should fit within the link capacity $C$.

*Characterizing an Optimal Solution.* We use the LCP tree for all addresses $\mathcal{BL} \cup \mathcal{G}$. Furthermore, to account for Eq.(21), we assign a cost, $t_p$, to every node in the LCP tree, representing the total traffic generated by prefix $p/l$, $t_p = g_p + b_p$.

*Proposition 3.4:* Given $\mathcal{BL}$, $\mathcal{G}$, $F_{max}$, and $C$, there exists an optimal solution of the FLOODING problem that can be represented as a pruned subtree of LCP($\mathcal{BL} \cup \mathcal{G}$), with the same root, up to $F_{max}$ leaves, and s.t. the total cost of the leaves be $\geq t_0 - C$.

*Proof:* The proof is along the same guideline of Prop.3.1. It can be shown that every feasible solution of FLOODING, $S$, can be mapped in another feasible solution, $S'$, which i) correspond to a subtree of LCP-tree($\mathcal{BL} \cup \mathcal{G}$) as described in Prop.3.4, and ii) whose collateral damage is smaller or equal to the collateral damage of $S$.

To see this, assume $S$ uses a prefix $\tilde{p}/\tilde{l}$, which is in not in LCP-tree($\mathcal{BL} \cup \mathcal{G}$). There cannot be good or bad sources in each of the two siblings prefixes, $\tilde{p}/(\tilde{l}+1)$. If this was the case, $\tilde{p}/\tilde{l}$ would be their longest common prefix, and consequently it would appear in LCP-tree($\mathcal{BL} \cup \mathcal{G}$).

Thus, there are two cases: If $\tilde{p}/\tilde{l}$ does not include any good or source we can simply remove it; otherwise we can filter only the branch that has some sources. Since the removed branch does not have active sources, the obtained solution is still feasible and the overall collateral damage is not increased (we are filtering a subset of what was already filtered). Iterating this process until all prefixes are in LCP-tree($\mathcal{BL} \cup \mathcal{G}$), prove that any feasible solution can be interpreted as a subtree of the LCP-tree, where the leaves are the actual filters used. Thus, also an optimal feasible solution can be represented in this way.

Finally, we have that, in order to have the allowed traffic within the capacity $C$, the filtered traffic, represented by the sum of costs $t_p$ at the subtree leaves, must be greater of equal than $t_0 - C$. ■

*Algorithm.* FLOODING is a 2-dimensional knapsack problem (2KP), with an additional capacity constraint, Eq.(22), that makes it harder. 2KP is a "very hard" problem: not only it is NP-Hard, but also the existence of a full polynomial time approximation scheme for this problem is unlikely to exist, since it would imply that $\mathcal{P} = \mathcal{NP}$ [6]. For FLOODING we obtain the following hardness result:

*Theorem 3.5:* The optimization problem FLOODING, in Eq.(19)-(22), is NP-Hard.

*Proof:* It is obvious that FLOODING is in $\mathcal{NP}$. To prove that it is also $\mathcal{NP}$-hard, we consider the KP problem with a cardinality constraint:

$$\max \sum_{i \in I} p_i x_i, \quad \text{s.t.} \sum_{i \in I} w_i x_i \leq C_1 \text{ and } \sum_{i \in I} x_i = k \tag{23}$$

which is known to be $\mathcal{NP}$-hard [4], and we show that it reduces to FLOODING. First, note that any solution of the above problem that uses $F < F_{max}$ filters can be transformed to another feasible solution with exactly $F_{max}$ filters, without increasing the collateral damage.[4] Therefore, the inequality in Eq.(20) can be replaced by an equality without affecting the collateral damage of the optimal solution. Second, we define $\bar{x}_{p/l} = 1 - x_{p/l}$, $\bar{F}_{max} = \Big(\sum_{p/l} 1\Big) - F_{max}$ and we rewrite the above problem:

$$\max \sum_{p/l} g_{p/l} \bar{x}_{p/l} \text{ s.t. :} \sum_{p/l} \bar{x}_{p/l} = \bar{F}_{max}, \tag{24}$$

$$\sum_{p/l} \Big(g_{p/l} + b_{p/l}\Big) \bar{x}_{p/l} \leq C, \quad \sum_{p/l: i \in p/l} \bar{x}_{p/l} \leq 1 \ \forall i \in \mathcal{BL} \tag{25}$$

For a given instance of Problem (23), we construct an equivalent instance of Problem (24)-(25) by introducing the following mapping. For $i = 1, \ldots, N$: $-\bar{g}_{ii} = p_i$, $(g_{ii} + b_{ii}) = w_i$. For $p/l$ that is not in the blacklist: $\bar{g}_{p/l} = 0$ and $(g_{p/l} + b_{p/l}) = C + 1$. Moreover, we assign $\bar{F}_{max} = k$ and $C = C_1$. With this assignment a solution to the KP problem (23) can be obtained by solving FLOODING and then taking the values of variables $x_{p/l}$ s.t $p/l$ is in the blacklist. ■

Therefore, we do not to look for a polynomial time algorithm. Instead, we designed a pseudo-polynomial time algorithm that optimally solves FLOODING, and whose complexity grows linearly with the number of active sources (either good or bad).

[4] This can be proved using the LCP-tree structure. Given a solution, $S$, with $F < F_{max}$ filters, (until $F < N$) there exist always a filter that can be replaced by two filters, corresponding its children. The solution constructed in such a way has $F + 1$ filters, keeps on blocking all IPs blocked in $S$, and has value less or equal than the value of $S$.

Let $z_p(F, c)$ be the minimum collateral damage solving FLOODING problem with $F$ filters and capacity $c$:

$$z_p(F, c) = \min_{\substack{n=0,\dots,F \\ m=0,\dots,c}} \{z_{s_l}(F-n, c-m) + z_{s_r}(n, m)\} \quad (26)$$

*Complexity.* The DP approach computes $O(CF_{max})$ entries for every node. Moreover, the computation of a single entry, given the entries of descendant nodes, require $O(CF_{max})$ operations, Eq.(26). We can leverage again the observation that we do not need to compute $CF_{max}$ entries for all nodes in the LCP tree. At a node $p$, it is sufficient to compute Eq.(26) only for $c = 0, ..., \tilde{C} = \min\{C, \sum_{i \in p/l} w_i\} \leq C$ and $f = 0, ..., \tilde{F}$. Therefore, the optimal solution to FLOODING, $z_{root}(F_{max}, C)$, can be computed in $O((N + |\mathcal{G}|)C^2)$ time. The algorithm has pseudo-polynomial complexity since it is polynomial in $C$ that cannot be bounded by the input length. More importantly, its complexity increases linearly with the number of IP sources in $\mathcal{BL} \cup \mathcal{G}$.

*FLOODING vs. BLOCK-SOME.* To see the connection between FLOODING and BLOCK-SOME, let us consider a partial Lagrangian relaxation of (19)-(22):

$$\max_{\lambda \geq 0} \Big\{ \min \sum_{p/l} \Big[(1-\lambda) g_{p/l} - \lambda b_{p/l}\Big] x_{p/l} + \quad (27)$$

$$+ \sum_{p/l} \lambda (g_{p/l} + \lambda b_{p/l}) - \lambda C \Big\}$$

$$\text{s.t.} \sum_{p/l} x_{p/l} \leq F_{max} \quad (28)$$

$$\sum_{p/l: i \in p/l} x_{p/l} \leq 1 \quad \forall i \in \mathcal{BL} \quad (29)$$

For every fixed $\lambda \geq 0$ problem (27)-(29) is equivalent to (19)-(22) for a specific assignments of weights $w_i$. This shows that dual feasible solutions of FLOODING are instances of BLOCK-SOME for a particular assignment of weights. The dual problem, in the variable $\lambda$, aims exactly at tuning the Lagrangian multiplier to find the best assignment of weights.[5]

### F. DIST(RIBUTED)-FLOODING

*Goal:* Consider a victim $V$ that connects to the Internet through its ISP and is flooded by a set of attackers (listed in a blacklist $\mathcal{BL}$), as in Fig.1(a). To reach the victim, attack traffic has to pass through one or more ISP routers; let $\mathcal{R}$ be the set of unique routers from some attacker to the victim. Let each router $u \in \mathcal{R}$ have capacity $C^{(u)}$ on the downstream link (towards $V$) and a limited number of filters $F_{max}^{(u)}$. We assume that the volume of good/bad traffic through every router is known. Our goal is to allocate filters across all routers, in a distributed way, so as to minimize the total collateral damage and avoid congestion on all links of the ISP network.

*Formulation.* Let the variables $x_{p/l}^{(u)} \in \{0, 1\}$ indicate whether or not filter $p/l$ is used at router $u$. Then the distributed filtering problem can be stated as:

$$\min \sum_{u \in \mathcal{R}} \sum_{p/l} g_{p/l}^{(u)} x_{p/l}^{(u)} \quad (32)$$

$$\text{s.t.} \sum_{p/l} x_{p/l}^{(u)} \leq F_{max}^{(u)} \quad \forall u \in \mathcal{R} \quad (33)$$

$$\sum_{p/l} \Big(g_{p/l}^{(u)} + b_{p/l}^{(u)}\Big)(1 - x_{p/l}^{(u)}) \leq C^{(u)} \quad \forall u \in \mathcal{R} \quad (34)$$

$$\sum_{u \in \mathcal{R}} \sum_{p/l \ni i} x_{p/l}^{(u)} \leq 1 \quad \forall i \in \mathcal{BL} \quad (35)$$

*Characterizing an Optimal Solution.* Given the sets $\mathcal{BL}$, $\mathcal{G}$, $\mathcal{R}$, and $F_{max}^{(u)}$, $C^{(u)}$ at each router, we have:

*Proposition 3.6:* There exists an optimal solution of DIST-FLOODING that can be represented as a set of $|\mathcal{R}|$ different pruned subtrees of the LCP-tree($\mathcal{BL} \cup \mathcal{G}$), each corresponding to a feasible solution of FLOODING for the same input, and s.t. every subtree leaf is not a node of another subtree.

*Proof.* Feasible solutions of DIST-FLOODING allocate filters on different routers s.t. Eq.(33) and (34) are satisfied independently at every router. In the LCP tree, this means having $|\mathcal{R}|$ subtrees, one for every router, each having at most $F_{max}^{(u)}$ leaves and associated blocked traffic $\geq t_0^{(u)} - C^{(u)}$, where $t_0^{(u)}$ is the total incoming traffic at router $u$. Each subtree on its own can be thought as a feasible solution of a FLOODING problem. Eq.(35) ensures that the same address is not filtered multiple times at different routers, to avoid redundant waste of filters. In the LCP-tree, this translates into every leaf of the different subtree appearing at most in one subtree. ■

*Algorithm.* Constraint (35), which imposes that different routers do not block the same prefixes, prevents us from a direct decomposition of the problem. To decouple the problem, consider the following partial Lagrangian relaxation:

$$L(x, \lambda) = \sum_{u \in \mathcal{R}} \sum_{p/l} g_{p/l}^{(u)} x_{p/l}^{(u)} + \sum_{i \in \mathcal{BL}} \lambda_i \Big(\sum_{u \in \mathcal{R}} \sum_{p/l \ni i} x_{p/l}^{(u)} - 1\Big)$$

$$= \sum_{u \in \mathcal{R}} \Big(\sum_{p/l} \Big(g_{p/l}^{(u)} + \lambda_{p/l}\Big) x_{p/l}^{(u)}\Big) - \sum_{i \in \mathcal{BL}} \lambda_i \quad (36)$$

where $\lambda_i$ is the Lagrangian multiplier (price) for the constraint in Eq.(35), and $\lambda_{p/l} = \sum_{i \in p/l} \lambda_i$ is the price associated with prefix $p/l$. With this relaxation, both the objective function and the other constraints immediately decompose in $|\mathcal{R}|$

[5]Problem (27)-(29) can be solved in a standard way with a projected subgradient method [4]

$$x_{p/l}^{(k)} = x_{p/l}^{*}(\lambda^{(k)}), \ \forall p, l \quad (30)$$

$$\lambda^{(k+1)} = \big[\lambda^{(k)} + \alpha_k \big((f_{p/l}^{g} + f_{p/l}^{b})(1 - x_{p/l}^{(k)}) - C\big)\big]^{+} \quad (31)$$

where, $x_{p/l}^{(k)}$ is the $k$th iteratation, $x_{p/l}^{*}(\lambda^{(k)})$ is the optimal solution of (27)-(29) for $\lambda = \lambda^{(k)}$, $\alpha_k > 0$ is the $k$th step size, and $[\cdot]^{+}$ indicates the projection over the set of non-negative numbers.

independent sub-problems, one per router $u$:

$$\min \sum_{p/l} \left(g_{p/l}^{(u)} + \lambda_{p/l}\right) x_{p/l}^{(u)} \tag{37}$$

$$\text{s.t.} \sum_{p/l} x_{p/l}^{(u)} \leq F_{max}^{(u)} \tag{38}$$

$$\sum_{p/l} \left(g_{p/l}^{(u)} + b_{p/l}^{(u)}\right)(1 - x_{p/l}^{(u)}) \leq C^{(u)} \tag{39}$$

The dual problem is:

$$\max_{\lambda_i \geq 0} \sum_{u \in \mathcal{R}} h_u(\lambda) - \sum_{i \in \mathcal{BL}} \lambda_i \tag{40}$$

where $h_u(\lambda)$ is the optimal solution of (37)-(39) for a given $\lambda$. Given the prices $\lambda_i$, every sub-problem (37)-(39) can be solved independently and optimally by router $u$ using e.g. Eq. (26). Problem (40) can be solved using a projected subgradient method, similarly to Eq.(30)-(31), as discussed in [4]. Note, however, that since $x \in \{0, 1\}$ the dual problem is not always guaranteed to converge to a primal feasible solution [7], [8].

*Distributed vs. Centralized Solution.* The above formulation lends itself naturally to a distributed implementation. Each router needs to only solve their own subproblem (37)-(39) independently from the others. A single machine (e.g. the victim's gateway or a dedicated node) should solve the master problem (40) to iteratively find the prices that coordinate all subproblems. Thus, at every iteration of the subgradient, the new $\lambda_i$'s need to be broadcasted to all routers. Given the $\lambda_i$'s, the routes independently solve a sub-problem each and return the computed $x_{p/l}^{(u)}$ to the node in charge of the master problem. Even in a centralized setting, our distributed scheme is efficient because it lends itself to parallel computation of Eq.(32)-(34).

## IV. Practical Evaluation

The focus of this paper is the design of optimal and computationally efficient algorithms for a variety of filter selection problems. In this section, we use real blacklists to demonstrate that filter optimization brings significant gain in practice. The reason is that, in practice, malicious sources appear clustered in the IP address space, a feature that is exploited by our algorithms. Due to lack of space, the simulations presented in this section are not exhaustive. However, they demonstrate the above point as well as some of the structural properties of the solution for BLOCK-ALL and BLOCK-SOME, which are at the heart of this framework. As discussed in section III, FLOODING is essentially an instance of BLOCK-SOME for a particular assignment of weights and DIST-FLOODING consists of several FLOODING problems.

### A. Simulation Setup

We analyzed 61-days traces from *Dshield.org* [3] - a repository of firewall and intrusion detection logs from about 2,000 different organizations. The dataset includes 758,698,491 attack reports, from 32,950,391 different IP sources. Each report includes, among other things, the malicious source IP and the victim's destination IP. By studying these logs, we verified that malicious sources are clustered in a few prefixes, rather than uniformly distributed over the IP space, which has also been observed by others [9]. This is an important observation in practice, because clustering in a blacklist means that a small number of filters is sufficient to block most malicious IPs at low collateral damage.

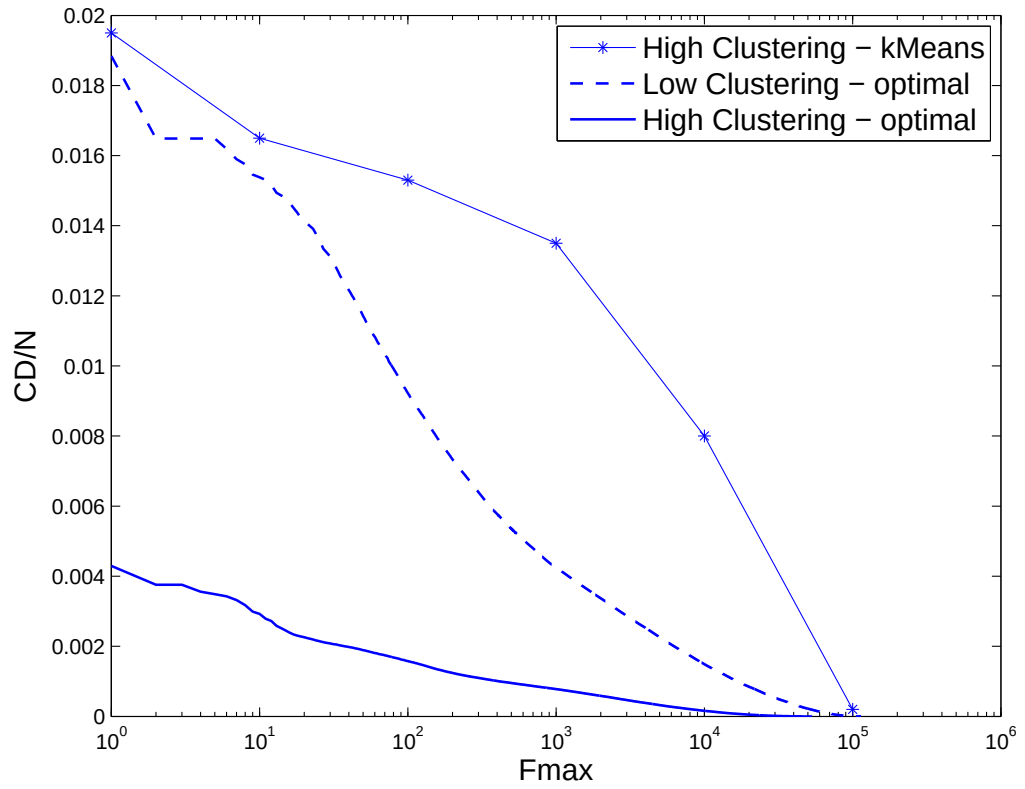

Fig. 4. BLOCK-ALL: collateral damage (CD) normalized over the number of malicious sources $N$) vs. number of filters $F_{max}$. We compare Algorithm 1 to K-means. (In particular, we simulated Lloyd's heuristic for K-means, which is NP-hard; we ran 50 runs to avoid local minima.) We also run Algorithm 1 on two traces, those with the highest and lowest degree of clustering.

We looked at each victim (individual IP destination) in the dataset; the set of sources attacking each victim is a blacklist for our simulations. This "view" varies considerably among victims. We also generated good traffic according to a realistic scenario: a domain hosting 20 servers, each server with average rate of 1,000 incoming good connections per second, each connection generating 5KB of traffic. We generated the good IP addresses according to the multifractal distribution in [10].

### B. Simulation Results

*BLOCK-ALL.* In Fig. 4, we chose two different victims, each attacked by large number (up to 100,000) of malicious IPs in a single day. We picked these particular two because they have the highest and the lowest degree of attack source clustering observed in the entire dataset. We ran Algorithm 1 on these two blacklists and made several observations. First, the optimal algorithm performs significantly better than a generic clustering algorithm that does not exploit the structure of IP prefixes. In particular, it reduces the collateral damage (CD) by up to 85% compared to K-means, when run on the same (high-clustering) blacklist. Second, as expected, the degree of clustering in a blacklist matters. The CD is lowest (highest) in the blacklist with highest (lowest) degree of clustering, respectively. Results obtained for other victim destinations and days were similar and lied in between the two extremes. A few thousands of filters were sufficient to significantly reduce collateral damage (CD) in all cases.

*BLOCK-SOME.* In Fig.5, we focus on the blacklist with the least clustering and thus the highest CD (dashed line in Fig.4). In this worst case scenario, an alternative to BLOCK-ALL is

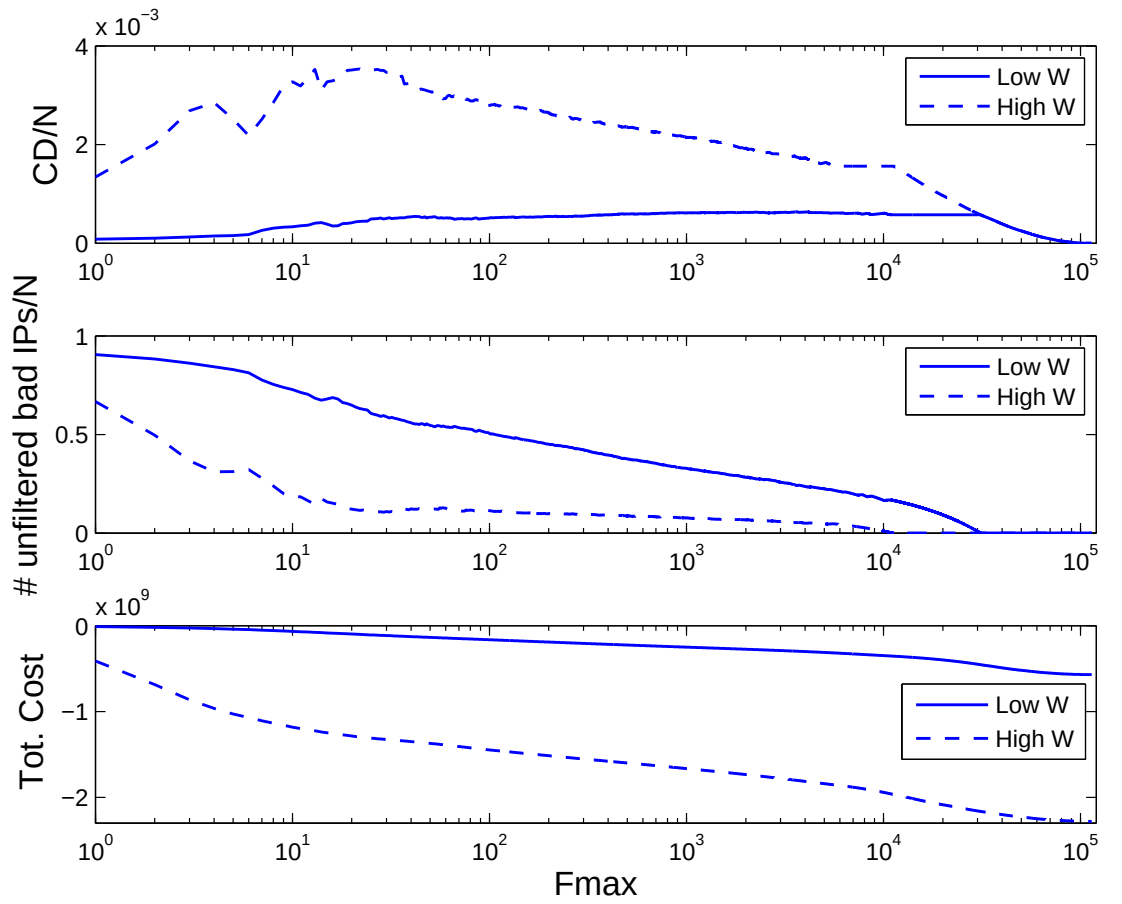

Fig. 5. BLOCK-SOME. (a) collateral damage (CD) (b) number of unblocked bad IPs (UBIP) (c) total cost ($CD - W \cdot UBIP$). The operator expresses relative tolerance to UBIP vs. CD by tuning the weight $W = \frac{w_b}{w_a}$. We considered a higher ($2^{14}$) and a lower ($2^{10}$) value of $W$.

BLOCK-SOME, which allows the operator to trade-off lower CD for unblocked bad IPs (UBIP) by appropriately tuning the weights. For simplicity, in Fig.5, we assigned the same weights $w_g$ and $w_b$ to all good and bad sources; however, the framework has the flexibility to assign weights to specific IPs. In Fig.5(a), the CD is always smaller than the corresponding CD in Fig.4; they become equal only when we block all bad IPs. In Fig.5(b), we see that BLOCK-SOME reduces the CD by 60% compared to BLOCK-ALL while leaving unblocked only 10% of bad IPs and using only a few hundreds a filters,

In Fig.5(c), the total cost decreases as $F_{max}$ increases. As defined in Eq.(12), this is the weighted sum of CD and UBIP. However, the behavior of these two competing factors is more complicated and depends strongly on the input blacklist. In the data we analyzed we observed that CD tends to first increase and then decrease with $F_{max}$, while UBIP tends to decrease.[6] The ratio $w_b/w_g$ captures the effort made by BLOCK-SOME to block all bad IPs and become similar to BLOCK-ALL.[7]

## V. Our Work in Perspective

### A. The Bigger Picture of Defense against Malicious Traffic

Dealing with malicious traffic is a hard problem that requires the cooperation of several components. In this paper, we did not propose a novel solution; instead, we optimized the use of filtering - a mechanism that already exists on the Internet today and is a necessary building block of any bigger solution. We focused on the optimal construction of filtering rules, which can be then installed and propagated by filtering protocols [11], [12]. We rely on a detection module, e.g. an intrusion detection system or historical data, to distinguish good from bad traffic and provide us with a blacklist. Detection is a difficult but orthogonal problem to the contribution of this paper. The sources of legitimate traffic are also assumed known, for estimating the collateral damage. Finally, we consider addresses in the blacklist to be true and not spoofed. This is reasonable today that attackers have the luxury to use botnets, and control a huge number of infected hosts for a short period of time, so that they do not even need to use spoofing. On 2005, less than 20% of addresses were spoofable [13], while in 2008, only 7% of addresses in Dshield logs were found likely spoofed [9]. Even if there is some amount of spoofed traffic, our algorithms treat it as the rest of malicious traffic and weight the cost vs. the benefit of blocking a source prefix (which may include both malicious spoofed and legitimate traffic). Looking into the future, there is also a number of proposals promising to enforce source accountability, including ingress filtering [14], self-certifying addresses [15], packet passports [16]. To the extent that spoofing interferes with the ability to define blacklists, our algorithms work best together with an anti-spoofing mechanism, but also do the best that can be done today without it.

A practical deployment scenario is that of *a single network under the same administrative authority*, such as an ISP or campus network. The operator can use our algorithms to create filtering rules, at a single or at several routers, in order to optimize the use of its own resources and defend against an attack in a cost-efficient way. Our distributed algorithm may also prove useful, not only for a distributed protocol of routers within the same ISP, but also in the future, when different ISPs start cooperating against common enemies. In a different context, our algorithms may also be applicable to *configure firewall rules to protect public-access networks*, such as university campus networks or web-hosting networks; although firewalls are implemented in software, there is still an incentive to minimize the number of their rules for performance reasons.

The following papers are related to our work. In [17], source filtering via ACLs was studied against DDoS attacks; however, the filters were heuristically selected and the approach was entirely simulation-based. There is a body of work *on firewall rule configuration* [18], which focuses on management and misconfigurations, not on resource allocation. Furthermore, they consider firewalls for enterprises, which are not supposed to be accessed from outside and thus can be protected without filtering rules. In our workshop paper [19], we also studied optimal source-based filtering by aggregating source addresses into continuous *ranges* (of numbers in $[0, 2^{32}-1]$) *not prefixes*. This was an easier problem that allowed for greedy solutions. Unfortunately, ranges are not implementable in ACLs; furthermore, it is well-known that ranges cannot be efficiently approximated by a combination of prefixes [5] . Therefore, despite the intuition we gained in [19], we had to solve the problem of prefix filtering from scratch in this paper.

[6]We can explain this as follows. When a new filter is available, the new optimal solution can be constructed by (i) blocking a new cluster of bad IPs (ii) splitting a blocked cluster into two filters or (iii) a combination of (i)&(ii)& merging of existing filters. For small $F_{max}$, option (i) is dominant: the inherent clustering allows to find a cluster that is not blocked yet; this increases CD and reduces UBIP. When this is not possible, option (ii) becomes dominant, CD decreases and UBIP remains constant or decreases slowly.

[7]Since we picked a ratio $w_b/w_g > 1$, bad IPs are more important. When $F_{max}$ is high, the algorithm first tries to cover small clusters or single bad IPs. In the case of high $W$, this happens around $10,000$ filters: CD remains almost constant in this phase, at the end of which all bad IPs are filtered (as in Fig.5(b)). In the final phase, the algorithm releases single good IPs, which are less important and all bad IPs are blocked similarly to BLOCK-ALL.

### B. Relation to Knapsack Problems

The optimal filter selection belongs to the family of *multi-dimensional knapsack problems (dKP)* [4]. The general dKP problem is well-known to be NP-hard. The most relevant variation to us is the *knapsack with cardinality constraint (1.5KP)* [21], [22], which has $d = 2$ constraints, one of them being a limit on the number of items: $\sum_{j \in \mathcal{N}} w_j x_j \leq C, \sum_{j \in \mathcal{N}} x_j \leq k$. The 1.5KP problem is also NP-hard.

These classic problems do not consider correlation between items. However, in filtering, the selection of an item (prefix) voids the possibility to select other items (all overlapping prefixes). dKP problems with *correlation between items* have been studied in [23], [24], where the items were partitioned into classes and up to one item per class was picked. In our case, a class is the set of all prefixes covering a certain address. Each item (prefix) can belong simultaneously to any number of classes, from one class (/32 address) to all classes (/0 prefix). To the best of our knowledge, we are the first to tackle the case where the items belong to classes that are *not a partition* of the set of items.

Finally, *continuous relaxations* do not help. Allowing $x_{p/l}$ to be fractional corresponds to rate-limiting of prefix $p/l$. However, there is no advantage neither from a practical (rate limiters are more expensive than ACLs, because in addition to looking up packets in TCAM, they also require rate and computation on the fast path) nor from a theoretical point of view (the continuous 1.5KP is still NP-hard [25].)

In summary, the special structure of filtering problems, i.e. the hierarchy and overlap of candidate prefixes, leads to novel variations of dKP that could not be solved by existing methods.

## VI. Conclusion

In this paper, we introduced a formal framework to study filtering problems. The framework is rooted at the theory of the knapsack problem and provides a novel extension of it. Within it, we formulated five practical problems, presented in increasing order of complexity. For each problem, we designed optimal algorithms that are also low-complexity (linear in the input size) in practical scenarios. We also highlighted connections between different problems: at the heart of all problems lies BLOCK-SOME; BLOCK-ALL and FLOODING are special instances for specific assignment of weights, and DIST-FLOODING decomposes into several independent FLOODING problems. Finally, we did simulations using Dshield traces; a key insight was that our algorithms can exploit the spatial clustering that is inherent in real blacklists.

There are several directions for future work. We plan to extend the framework to dynamically update the filtering rules as blacklists change over time, combine source- with destination-based filtering, deal with adversarial scenarios, and study the interaction between filtering and detection mechanisms. We will also provide a more extensive experimental evaluation, which is not the focus of this paper.